\documentclass[twocolumn]{revtex4}
\usepackage{graphicx}

\begin{document}

\title{Implementing Elements of \emph{The Physics Suite} at a Large Metropolitan Research University}
\author{Costas Efthimiou$^{1}$, Dan Maronde$^{1}$, Tim McGreevy$^{2*}$\thanks{},
Enrique del Barco$^{1}$, Stefanie McCole$^{3\dagger}$} \affiliation{$^{1}$Department of Physics,
University of Central Florida, Orlando, FL.}
\affiliation{$^{2}$Department of Physics,
Embry-Riddle Aeronautical University, Daytona Beach, FL.}
\affiliation{$^{3}$Department of Physics, Old Dominion University,
Norfolk, VA.}

\begin{abstract}
A key question in physics education is the effectiveness of the
teaching methods.  A curriculum that has been investigated at the
University of Central Florida (UCF) over the last two years is the
use of particular elements of \emph{The Physics Suite}. Select sections of
the introductory physics classes at UCF have made use of
\emph{Interactive Lecture Demonstrations} as part of the lecture
component of the class. The lab component of the class has
implemented the \emph{RealTime Physics} curriculum, again in
select sections. The remaining sections have continued with the
teaching methods traditionally used. Using pre- and post-semester
concept inventory tests,  
a student survey,  student interviews,
and a standard for successful completion of the course, 
 the preliminary data indicates
improved student learning.
\end{abstract}
\maketitle

{\renewcommand{\thefootnote}
    {\fnsymbol{footnote}}
    \footnotetext[1]{Prior to August 2008:{\itshape Department of Physics, University of Central Florida, Orlando, FL.}}
    \footnotetext[2]{Prior to August 2009:{\itshape Department of Physics, McDaniel College, Westminster, MD.}}
}

\section{Introduction}
The University of Central Florida (UCF) is a large metropolitan
university located in Orlando, Florida. Shortly before this study was started, the U.S. Department of
Education National Center for Education Statistics ranked UCF as
the sixth largest university in the country based on a Fall 2006
enrollment of 46,646 students \cite{USDE}. Since that time, the
university's enrollment has grown to over 55,000, making UCF the
third largest university in the United States going into the Fall
2009 semester \cite{WFTV,UCF1}.
 The calculus-based
introductory physics courses, Physics I (Newtonian Mechanics) and
Physics II (Electricity and Magnetism with some Optics) are
required for all students of the College of Engineering and
Computer Science. This college alone had an enrollment of over
6000 in the spring of 2009 \cite{ECS}. The large number of
students going through the introductory classes each semester,
along with the students from the College of Sciences, which
includes the Physics Department, and a large contingent of
students from other colleges taking the department's physical
science class make UCF's Physics Department the second-largest in
the country based on the number of credit-hours taught per
semester \cite{AIP}. The teaching of these introductory classes is
a large-scale operation, with ten to twelve lecture sections of
eighty to over one-hundred students taught each semester, along
with twenty to twenty-two lab sections of thirty to thirty-five
students each. Retention of the students within the majors that
make up the science, technology, engineering and mathematics
fields (STEM) is a challenge for UCF as well as universities
nationally \cite{NSF}. The introductory physics courses are one of
the first large hurdles that students must clear on their way to a
STEM degree.

There are many innovative approaches to teaching introductory
physics (see Ref. \cite{Redish} and references therein). One that has been demonstrated to effectively improve
students' conceptual learning is curriculum presented in \emph{The
Physics Suite} \cite{Redish,PS}. Some elements of \emph{The Physics
Suite} apply readily to the large-scale approach to the teaching
of the introductory classes at UCF, with little or no change to
the scheduling format or the infrastructure used by the
department. \emph{Interactive Lecture Demonstrations} (ILD) \cite{ILD}
can be used effectively in a large lecture hall setting, so they
can be implemented in any of the sections scheduled at UCF.
Similarly, \emph{RealTime
Physics} (RTP) \cite{MOD1,MOD3,MOD4} is designed for use in a
lab setting with groups of two to four students performing
experiments using computers and probeware, much of which is
already used in the UCF introductory physics labs. These elements
of \emph{The Physics Suite} bring physics education research-based
curriculum into an existing large-scale physics instruction
program while maintaining the department's existing method of
scheduling and class structure and making use of the department's
existing equipment.

The evaluations outlined in this paper provide evidence of
improved student learning due to the effectiveness of including
\emph{The Physics Suite} methods as a major part of the
instruction in the introductory physics classes. The large scale
of UCF's introductory physics program has allowed a comparison of
conceptual learning between significant numbers of students taking
part in the new method of instruction with those in traditional
instruction, as well as those with partial involvement in both
methods, over a short span of time. The number of sections of each class available for this survey has also provided the opportunity to study the relative benefits of \emph{Interactive Lecture Demonstrations} and \emph{RealTime Physics} when implemented seperately as well as in concert.

\section{Introductory Physics classes at UCF, and changes made to implement \emph{The Physics Suite}}
The lecture sections of the introductory physics classes at UCF
are taught in an auditorium-style lecture hall seating 80-120
students. Traditional format lectures are taught in a visual method
determined by the instructor ---  
that is, 
either projected powerpoint  slides or flashing of transparencies or blackboard/whiteboard presentation. A few
instructors have made occasional use of electronic student
response systems (ERS) to incorporate some active student
participation into the format and/or have used a combination of the visual methods for the delivery
of the lectures.

The sections chosen for participation in \emph{The Physics Suite}
project used \emph{Interactive Lecture Demonstrations} 
during one lecture class each week. A demonstration relating to
the current topic being addressed in the class is set up. Students
are asked to make a prediction of what results will be seen during
the demonstration. The predictions are recorded as a poll, using
the ERS. Students are then encouraged to interact in small groups,
discussing their predictions and explaining their reasoning. After
the discussion interval, students are polled again and the results
are displayed. The demonstration is then conducted and students
can see the actual results. Finally, the physical principles
involved are discussed, with participation both from students who
had reasoned correctly and from some who had based their answers
on incorrect preconceptions. This approach of direct confrontation
of preconceptions with actual demonstrations, requiring the
students to thoroughly reason through the problem both before and
after the demonstration has been shown to be an effective teaching
method. It develops deeper understanding by connecting students'
intuition to the concepts addressed, and it is effective in
changing flawed preconceptions
\cite{CBZ,RRH,MCD,ST1}.

The lab component of the introductory classes is a separate, one
credit-hour class. The classrooms are arranged for eight stations
of four students each, and each station equipped with a computer
and Pasco CI-6560 interface. The experiments used in the
traditional format lab are mostly``cookbook''-style. The students
present their results in a report composed outside of class and
turned in the following meeting. The format of the report is
determined by the instructor. Part of the focus of the manual is
on experimental set-up and method, and there is some introduction
to error analysis.

The lab sections using \emph{RealTime Physics} as their lab
manual are set in the same rooms and make use of much of the same
equipment as the traditional format labs.  The RTP manual makes
use of a guided-inquiry approach to the execution of the
experiments. After a brief outline of the experimental set-up, the
students are asked to make predictions of the results they expect
to see. Like the ILDs in the lecture sections, this prediction
step will target preconceptions and lead to discussion of the
underlying principles among members of the lab group. For most
activities, the data is plotted in real time, as the experiment
takes place. The students are encouraged to make multiple runs of
each experiment, watching the relation between the physical
activity and the mathematical plot. The actual results are
compared with the predictions made earlier. Questions are asked to
get the students to think about the physical principles
determining the experimental results. Some mathematical analysis
is done with the graphs and data from the experiment to show
relations between physical quantities.

An example of an activity performed during one of the
one-dimensional motion labs from Physics I follows:  A student
holds a ball above a motion detector that is hooked to a computer.
The students in the group predict what will happen to velocity and
acceleration for the ascent, turning point, and descent of the ball thrown vertically.  Then the
ball is thrown in the air so the motion detector can plot velocity
versus time and acceleration versus time.  Now students discuss
the results and can see that even though velocity
changes, acceleration is constant. The plots of velocity and
acceleration are compared, illustrating the mathematical relation
between the quantities. Careful attention to the graphs also helps
illustrate the vector nature of the quantities \cite{MRZ,RJB,GA}.

The lab report is completed within the lab class time, consisting
of answering the questions posed during the activities. There is a set of 
homework questions following each lab, designed to reinforce concepts
developed through the activities in the lab. Through the Fall 2007
semester, the first semester of the project, the homework was
optional. Since then, the students have been required to turn this
in at the start of the following lab. The emphasis of the RTP labs
is on understanding the physical concepts addressed and
interpreting the graphs of mathematical relations followed by the
physical quantities \cite{ST2,RSS}.

\section{Method of evaluation}
To evaluate the effectiveness of these portions of \emph{The Physics Suite} method
at UCF, all the students in the introductory physics classes were
given a pre-test during the first meeting of the lab section, and
a post-test on the last day of the lab. The tests used are concept inventory-style
tests and the same tests were given to all students both times.
The Physics I test is the \emph{Force and Motion Concept
Evaluation} and is included as an appendix to the
Thornton/Sokoloff article cited \cite{LM}. The Physics II test is
the \emph{Electric Circuits Concept Evaluation}, also created by
Sokoloff and Thornton. The difference between each student's score
between the pre- and post-tests was not the score evaluated.
Instead, students' pre- and post-test scores were compared based
on the percentage of material that they learned during the
semester. The improvement score, normalized gain as introduced by Hake, calculates what
percentage of the questions missed on the pre-test are answered
correctly on the post-test. The formula used to calculate the
normalized gain is $\mbox{Gain}=\frac{(Post~Correct)-(Pre~Correct)}{(Perfect~Score)-(Pre~Correct)}\times 100$. \cite{RRH2}

The students in each semester registered for their lecture and lab
classes without knowing if either would include a component of
\emph{The Physics Suite}, so their placement was random. A small
number of students each semester take either the lab or lecture
component alone. The students taking only the lecture component
did not take the concept evaluation tests.

The scores from the pre- and post-tests were grouped into four
categories: students taking an ILD lecture section and an RTP lab,
students taking an ILD lecture section and a traditional format (TF) lab, students
taking a traditional format (TF) lecture and an RTP lab, students taking
a TF lecture and a TF lab. Students who missed
either test due to add/drop of a class or an absence on either
test day were not counted.  Also, for any student who had taken a
pre-test and switched sections and took it again the lower score
was kept. These statistics were used to evaluate the effect of
\emph{The Physics Suite} elements on conceptual learning.

The students' successful completion of the course, defined as
attaining a ``C'' or better, is compared among the four
categories. The percentage of each category that successfully
completed the course during the study is presented for both
Physics I and Physics II.

Finally, student opinions were solicited. Interviews were
conducted with student volunteers after two semesters, the Spring
2008 semester and the Spring 2009 semester. The students did not receive any monetary
or grade compensation for the interviews. In the spring of 2008,
twenty-eight students were interviewed, eight students from
Physics I and twenty from Physics II.  Six of the students were
Physics II RTP students who had taken Physics I during the fall
with the RTP method. A survey was given to all sections of the lab
classes at the end of the Spring 2009 semester. Excerpts from the
 interviews and results from the  survey are presented.

\section{Evaluation: Concept Test Results}
The students' conceptual learning as measured by the performance
on the \emph{Force and Motion Concept Evaluation}(FMCE) and
\emph{Electric Circuits Concept Evaluation} (ECCE) pre- and
post-tests is presented first. The comparison is made between
students taking sections using either \emph{Interactive Lecture
Demonstrations}(ILD) in their lecture, \emph{RealTime
Physics}(RTP) in their lab, or both components of \emph{The
Physics Suite}(ILD \& RTP) and those taking only traditional
format classes(TF) in both their lecture and lab component.

\begin{figure}
    \centering
        \includegraphics[width=0.50\textwidth]{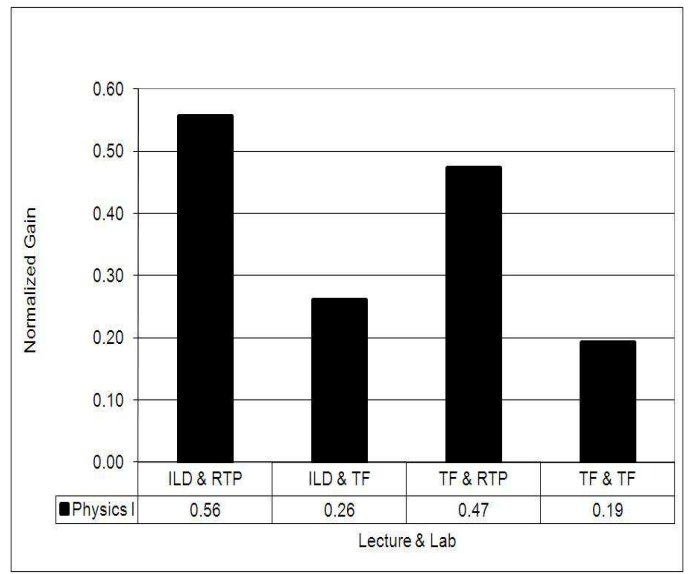}
    \caption{Cumulative Physics I results}
    \label{fig:2048cumulative}
\end{figure}
\begin{figure}
    \centering
        \includegraphics[width=0.50\textwidth]{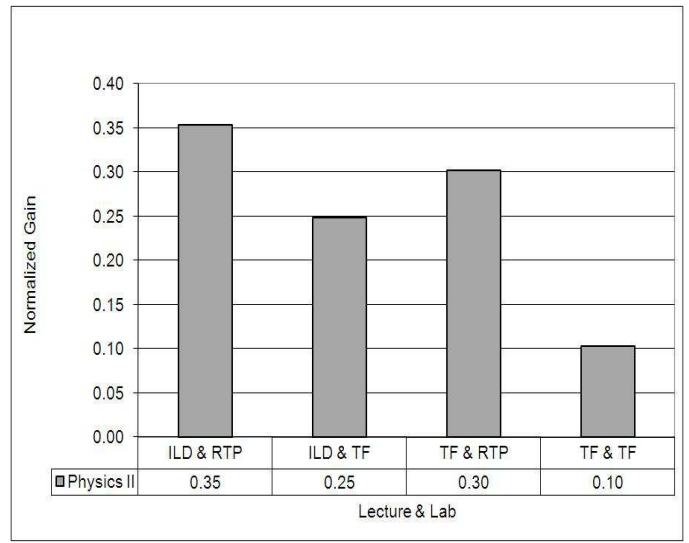}
    \caption{Cumulative Physics II results}
    \label{fig:2049cumulative}
\end{figure}
Figures \ref{fig:2048cumulative} and \ref{fig:2049cumulative} show
the two-year cumulative performance in Physics I (PHY2048) and
Physics II (PHY2049) separately. The gains for all formats are
higher in Physics I, with students taking both ILD and RTP
components gaining over 50\% compared to 19\% for students having
both components in the traditional format. In Physics II, the
gains are less for all formats, but students taking any component
of \emph{The Physics Suite} more than doubled their gain over the
TF-only group.

The remaining figures for this section show the performance of
students in the introductory classes by semester for the first two
years of the project, from the most recent results back to the
start of project. Note that in PHY2049 there are no students in
the TF lab categories after the first semester of implementation.
After positive results in the initial semester of the project, all
sections of the Physics II lab component were switched to the RTP
format.
\begin{figure}
    \centering
        \includegraphics[width=0.50\textwidth]{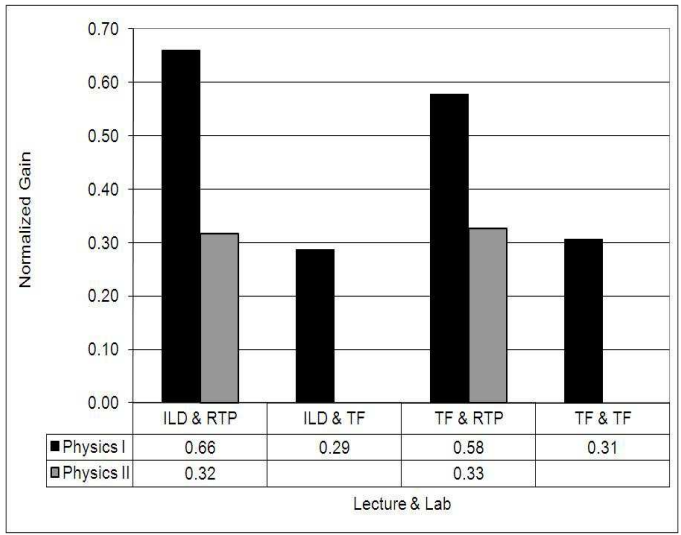}
    \caption{FMCE \& EECE results for Physics I \& II from Spring 2009}
    \label{fig:sp2009}
\end{figure}
\begin{figure}
    \centering
        \includegraphics[width=0.50\textwidth]{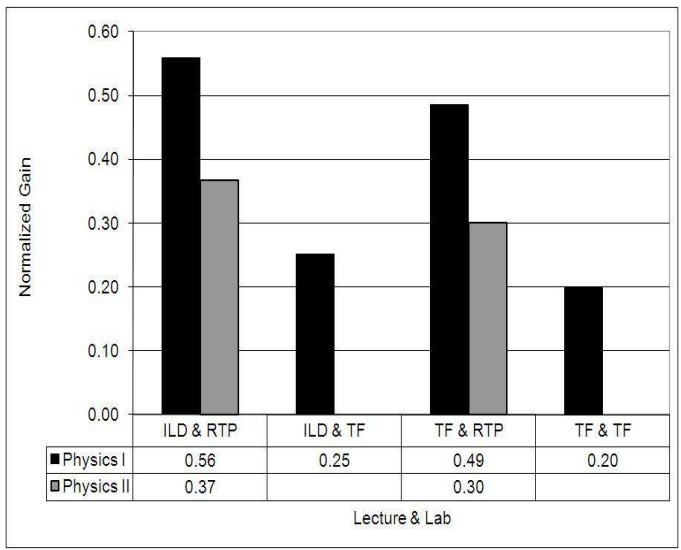}
    \caption{FMCE \& EECE results for Physics I \& II from Fall 2008}
    \label{fig:f2008}
\end{figure}
\begin{figure}
    \centering
        \includegraphics[width=0.50\textwidth]{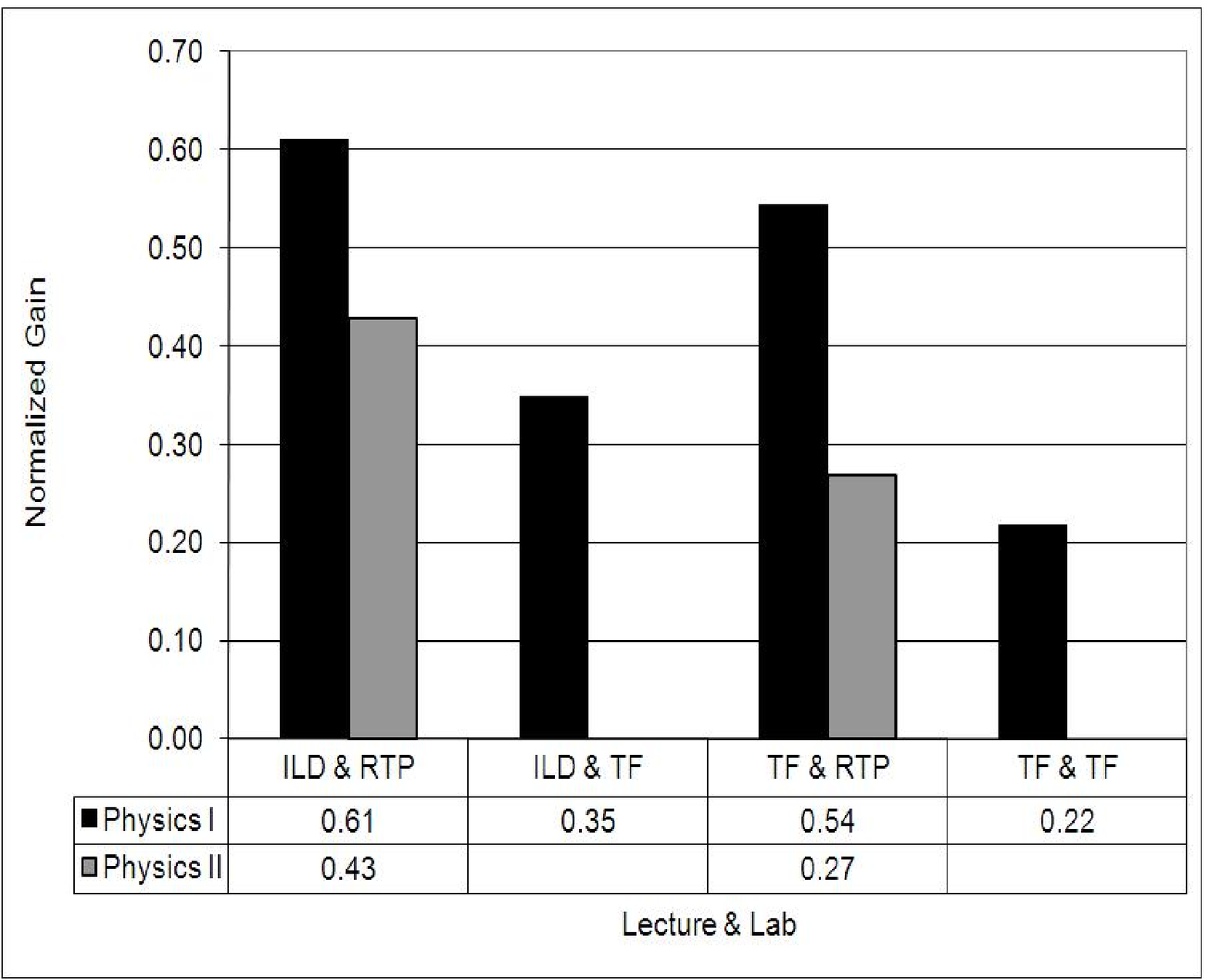}
    \caption{FMCE \& EECE results for Physics I \& II from Spring 2008}
    \label{fig:sp2008}
\end{figure}
\begin{figure}
    \centering
        \includegraphics[width=0.50\textwidth]{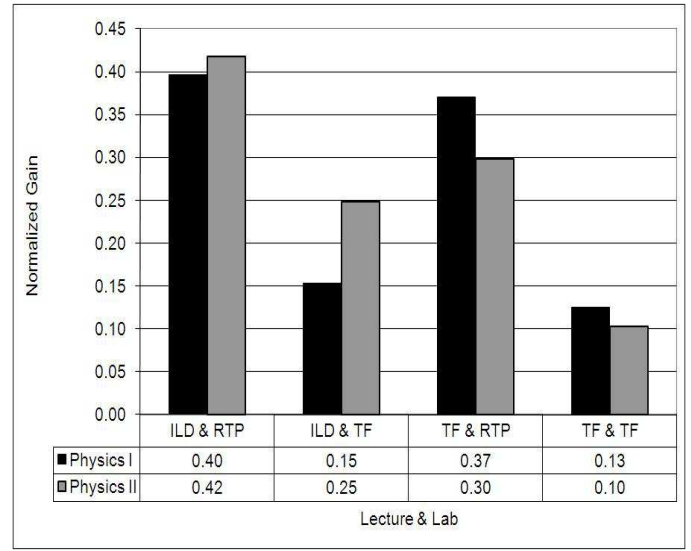}
    \caption{FMCE \& EECE results for Physics I \& II from Fall 2007. This was the only semester to include TF labs in Physics II.}
    \label{fig:f2007}
\end{figure}

The latest results from the Spring semester of 2009 are shown in
Figure \ref{fig:sp2009}. Figure \ref{fig:f2008} shows the results
from the Fall semester of 2008. Figure \ref{fig:sp2008} shows the
performance in the Spring of 2008. Finally, Figure \ref{fig:f2007}
shows the Fall of 2007, the first semester of implementation.

For Physics I, the number of students in each category and for
each semester is given in Table \ref{physics1_numbers}. The Physics II numbers are in Table \ref{physics2_numbers}. Fall 2007 was the only semester with traditional format labs in Physics II. 

\begin{table}[htdp]
\centering
\large{{\bf Physics I Numbers by Semester}}
\begin{tabular}{lcccc} 

\small{Semester} & \small{ILD\&RTP} & \small{ILD\&TF} & \small{TF\&RTP} & \small{TF\&TF}\\
\hline 

Fall 2007 & 26 & 57 & 72 & 270 \\
Spring 2008 & 15 & 42 & 43 & 186 \\
Fall 2008 & 19 & 49 & 60 & 104 \\
Spring 2009 & 32 & 121 & 34 & 120 \\
\hline
Total & 92 & 269 & 209 & 680

\end{tabular}
\caption[Physics I Numbers by Semester]{The number of students taking the FMCE, by semester.}
\label{physics1_numbers} 
\end{table}

\begin{table}[htdp]
\centering
\large{{\bf Physics II Numbers by Semester}}
\begin{tabular}{lcccc} 

\small{Semester} & \small{ILD\&RTP} & \small{ILD\&TF} & \small{TF\&RTP} & \small{TF\&TF}\\
\hline 

Fall 2007 & 17 & 21 & 66 & 191 \\
Spring 2008 & 15 &  & 141 &  \\
Fall 2008 & 46 &  & 210 &  \\
Spring 2009 & 78 &  & 210 &  \\
\hline
Total & 156 & 21 & 627 & 191

\end{tabular}
\caption[Physics II Numbers by Semester]{The number of students taking the EECE, by semester.}
\label{physics2_numbers} 
\end{table}
\section{Evaluation: Successful Class Completion}

The FMCE and ECCE test conceptual understanding. Another important
measure of the success of a course is given by the percentage of
students that successfully complete the course. In the case of the
introductory physics courses, the majority of the students are
from the College of Engineering and Computer Science, which
requires a C or better in the core courses.
\begin{figure}
    \centering
        \includegraphics[width=0.50\textwidth]{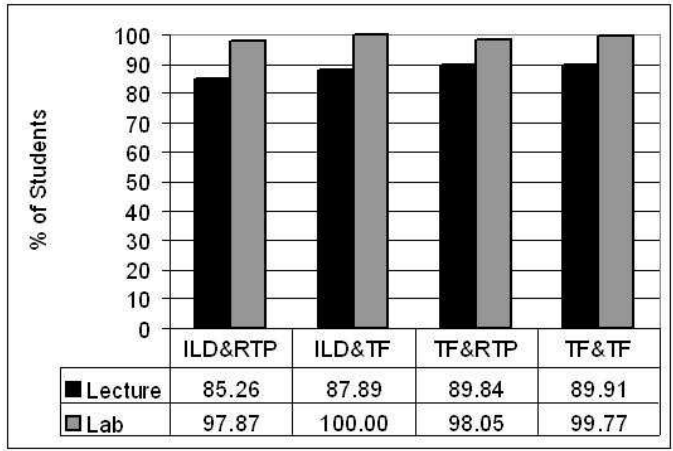}
    \caption{Percentage of students successfully completing the introductory physics classes (Fall 2007-Spring 2009)}
    \label{fig:abc}
\end{figure}
Figure \ref{fig:abc} shows the percentage of students attaining at
least a C grade in Physics I from the Fall 2007 semester through
the Spring 2009 semester. Again, the performance is divided into
the categories based on the combination of lecture and lab
instruction method. This is the one area where taking both ILD and
RTP components has not shown to be an advantage.
The effect may be masked by less grade inflation in the ILD classes taught by the project's PI(C.E.), and by the stricter rules of student evaluation he has established in the RTP labs. 
Since the combination of the two elements of \emph{The Physics Suite} created the largest gain measured in conceptual
understanding, it may be that it leads to a bigger split between
the successful students and those that are still struggling. This
may lead to a higher withdrawal rate. 
At this point the
investigation has not differentiated between withdrawals and
failures. The retention issue is an important one and the other
apparent successes of the project make it seem that it should help
with this also.

\section{Evaluation: Student Input}

The evaluation also involved student input. A survey was given to
all students taking the post-test at the end of the Spring 2009
semester. The survey consisted of fifteen statements relating to
the benefits of the lab component of the course, the correlation
between the lab and lecture components, the recitation sessions
conducted during the lab meeting time, and possible ways to
maximize the benefit of the lab component. Students were asked to
rank their agreement with each statement on a five-point scale:
strongly agree, agree, neutral, disagree, or strongly disagree.

Here the responses of Physics I students are shown for two
statements that are key to both the motivation for, and the
success of the implementation of \emph{The Physics Suite}. They
refer to the goal of challenging common preconceptions held by
students of the introductory physics classes, which \emph{The
Physics Suite} curriculum is written specifically to address, and
the goal of better coordinating the lab component topics with the
lecture component topics chronologically during the semester.
\begin{figure}
    \centering
        \includegraphics[width=0.50\textwidth]{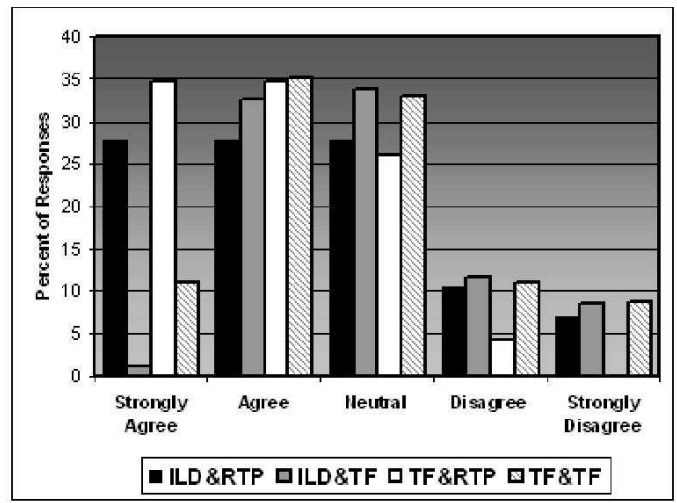}
    \caption{Percentage of student responses to the statement: ``The experiments performed in the lab have caused me to 
               change my mind about how some physical process works.''}
    \label{fig:survey q8 2048}
\end{figure}

Figure \ref{fig:survey q8 2048} shows the responses to the
statement, ``The experiments performed in the lab have caused me
to change my mind about how some physical process works.'' The
students who had taken a \emph{RealTime Physics} lab (RTP) had a
higher percentage of positive responses than those taking the
traditional format lab (TF). The strongest positive response came
from students taking the RTP lab, with the traditional lecture
component. These students seemed to have a strong sense that the
lab exercises contributed to their understanding of the physical
principles involved. The responses of the students taking an RTP
lab component with an ILD lecture, although still positive, were
not as strong as the RTP lab with traditional lecture students.
This is the group that performed the highest on both the
conceptual test and in the lecture component grade. The lower
positive response may be due to the wording of the statement.
These students may feel that their preconceptions were challenged
during the ILD lecture as well as during the RTP lab experiments.
\begin{figure}
    \centering
        \includegraphics[width=0.50\textwidth]{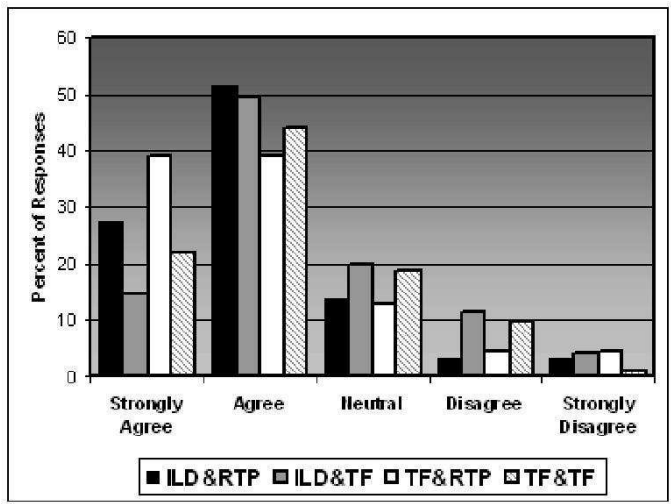}
    \caption{percentage of student responses to the statement: ``There is a connection between the concepts addressed by the 
               lab experiments and the theory covered in the lecture class.''}
    \label{fig:survey q2 2048}\end{figure}

Figure \ref{fig:survey q2 2048} shows the percentage of the
responses to the statement, ``There is a connection between the
concepts addressed by the lab experiments and the theory covered
in the lecture class.'' Again, students have responded more
positively if they had the RTP lab than if they had a TF lab.

Student volunteers were interviewed following both the Spring 2008
and Spring 2009 semesters. The earlier interviews have been
transcribed and the student feedback is summarized here.

On major items, such as `Do you like the format?', `Have the
experiments changed misconceptions?', and `Do the experiments help
you learn physics?', the majority of students gave positive
feedback.  The interactive nature of the RTP labs was praised by a
number of students. Many students claimed to have had
preconceptions modified by conducting the experiments in the RTP
lab. A question in the interview addressed the homework and most
students had a positive response. It was an indication the
homework was fulfilling its intended purpose. The interviews were
conducted in a focus group setting. Examples of two questions and
typical student responses follow. The responses have been
paraphrased to accommodate grouping of like responses. 
\begin{enumerate}
\item {\bfseries Do you like this (overall) format as compared to
a traditional lab format where you take all the data and later
write a report? Why?}
\begin{itemize}
\item [-]  Much more interactive 
\item [-]  More effective because you forget when you try and write the report at home 
\item [-] Somewhat, explains to a point that makes you find out a purpose 
\item [-] Good to answer questions as going through 
\item [-] Does more to reinforce
\end{itemize}
\item {\bfseries Do the lab report questions help understand the material?}
\begin{itemize}
\item [-] Helps understand 
\item [-] Makes you stop and think
\item [-] Helps if you take your time to do it 
\item [-] Observing then writing reinforces 
\item [-] Seeing the result really helps with concept
\end{itemize}
\end{enumerate}
\section{Implementation Issues}

Although the students are encouraged to do their best on the pre-
and post-tests, and most do, there are some who just quickly mark
down an answer. There needs to be an incentive for all to try
their best. It cannot be directly related to their grade because,
in the case of the pre-test, it would be unfair to grade students
on material they have not seen. The graduate teaching assistants
have been asked to offer some form of extra credit to students as
an incentive to give a full effort on both tests. Some methods
that have been tried are rewarding the top-scoring students on the
pre-test, rewarding both the top scoring and highest normalized
gain students on the post-test, and counting the post-test as a
full-score replacement quiz grade based on effort (as judged by
the instructor).

Another issue is that as time goes on students from the past RTP
labs will have corrected homework that they can share with
current RTP students. Although this would not show directly in the
pre/post test results, some students may not put full effort into
learning. A possibility may be to have different homework sets
appended to those in the lab manual.

Attaining a good correlation between the lab and the current topic
being covered in the lecture component is always a challenge and
was part of the motivation for undertaking the project. Much of
the problem involves the different pace of the various lecture
instructors. The RTP labs for Physics I have been in fairly good
agreement with the lectures. Physics II has been more difficult.
The labs start immediately with circuits, while the lecture deals
with static charges, electric force and electric fields. There is
also a point where the lecture starts addressing magnetism and
there are no corresponding labs. The geometric optics labs have
been inserted into the lab curriculum at this point, and then
circuits are resumed once the lecture begins with some
introduction of induction. Although students have had some
complaints about the disconnect between the two components, most
have been glad for the experience with circuits when that topic is
reached in lecture.

\section{Conclusion}
There is strong evidence, based on the results from the pre- to
post-test gains on the FMCE and ECCE, that \emph{The Physics
Suite} elements implemented into the introductory physics classes
at UCF have a positive effect on students' conceptual learning.
This is especially true when both the \emph{Interactive Lecture
Demonstrations} and \emph{RealTime Physics} labs are used in
conjunction. However, the investigation shows evidence that either component benefits student learning when used alone. 
This makes \emph{The Physics Suite} a good option for a large-scale physics program with many different instructors. 
\emph{RealTime Physics} as a stand-alone method for lab instruction pairs well with the lecture component of the course, 
whether or not the lecture instructor chooses to include \emph{Interactive Lecture Demonstrations} as part of the course.

This improved performance does not seem to have affected the
number of students who successfully complete the courses, although
those that do succeed are doing so at a higher level. The students
that withdraw or fail often do not attend class, or do not put any
effort into their studies of the subject. Even if the presentation
of the material is improved, it does not have an effect on this
group. The effort to stimulate learning in the lowest performing
students is an area that will require further focus.

There is also evidence that the students themselves feel that the
approach of \emph{The Physics Suite} has positively impacted their
learning. Responses to the survey questions and statements made
during the interviews indicate a perception among the students
that the inquiry-based activities of \emph{RealTime Physics} and
the active participation in lecture required in \emph{Interactive
Lecture Demonstrations} play a large role in helping them grasp
the difficult material of the introductory physics classes.

The success of the project through the first two years, based on
the evaluations done and also on positive feedback from the
students themselves indicates that there is good reason to
continue with, and expand, the implementation of \emph{The Physics
Suite} at The University of Central Florida.

\section{Acknowledgments}
The authors would like to extend thanks for the help provided by Professors David Sokoloff, Ronald Thornton and Priscilla Laws, who have been 
the consultants in the implementation of the \textit{The Physics Suite} at UCF, and Professor Ken Krane,
who has been the evaluator. The lab manuals and concept inventory tests used for Physics I and II have been created by David Sokoloff, 
Ronald Thornton and Priscilla Laws.
\cite {ILD,MOD1,MOD3,MOD4}.  

This work was supported by the Department of Undergraduate Education
in the National Science Foundation under Grant No. DUE 0633157.  S. McCole was supported by an REU grant of 
the National Science foundation (DMS 0649159). 

Any opinions, findings, and conclusions or recommendations expressed
in this material are those of the author(s) and do not necessarily
reflect the views of the National Science Foundation.


\begin{thebibliography}{99}

\bibitem {USDE} U.S. Department of Education National Center for Education Statistics, http://nces.ed.gov/fastfacts/
\bibitem {WFTV} WFTV, Orlando, FL, http://www.wftv.com/education/
\bibitem {UCF1} University of Central Florida, UCF Today, http://today.ucf.edu/ucf-moves-up-now-nations-3rd-largest/
\bibitem {ECS} University of Central Florida College of Engineering and Computer Science, http://www.cecs.ucf.edu/about/
\bibitem {AIP} American Institute of Physics, AIP Pub. No. R-394.14, http://www.aip.org/statistics/
\bibitem {NSF} National Science Foundation, http://www.nsf.gov/news/
\bibitem{Redish} E. F. Redish, \textit{Teaching Physics with the Physics Suite}, Wiley 2003. 
\bibitem {PS} Karen Cummings, Priscilla Laws, Edward F. Redish, Patrick Cooney, David Sokoloff, Ronald Thornton, {\it The Physics Suite}, John Wiley and Sons, Inc. 2004.
\bibitem {ILD} David Sokoloff, Ronald Thornton. {\it Interactive Lecture Demonstrations, Active Learning in Introductory Physics}, John Wiley and Sons, Inc. 2006.
\bibitem {LM} Ronald Thornton, David Sokoloff. {\it Assessing Student Learning of Newton's Laws: The Force and Motion Conceptual Evaluation and the Evaluation of Active Learning Laboratory and Lecture Curricula}, Am. J. Phys. 66(4). April 1998.
\bibitem {MOD1} David Sokoloff, Ronald Thornton, Priscilla Laws. {\it Real Time Physics Active Learning Laboratories: Module 1 Mechanics}, John Wiley and Sons, Inc. 2004.
\bibitem {MOD3} David Sokoloff, Ronald Thornton, Priscilla Laws. {\it Real Time Physics Active Learning Laboratories: Module 3 Electric Circuits}, John Wiley and Sons, Inc. 2004.
\bibitem {MOD4} David Sokoloff, Ronald Thornton, Priscilla Laws. {\it Real Time Physics Active Learning Laboratories: Module 4 Light and Optics}, John Wiley and Sons, Inc. 2004.
\bibitem {CBZ} J. Clement, D. Brown, A. Zeitsman, {\it Not all preconceptions are misconceptions: Finding ``anchoring conceptions'' for grounding instruction on students' intuition},  Int. J. Sci. Educ. 11 (spec. issue), 554-565 (1989).
\bibitem {RRH} R. R. Hake, {\it Socratic pedagogy in the introductory physics laboratory}, Phys. Teach. 33, 1-7 (1992).
\bibitem {RRH2} R. R. Hake, {\it Interactive engagement versus traditional methods: A six-thousand student survey of mechanics test data for introductory physics courses}, Am. J. Phys. 66(1). January 1998.
\bibitem {MCD} L. C. McDermott, P. S. Shaffer, {\it Research as a guide for curriculum development: An example from introductory electricity. Part I. Investigation of student understanding},  Am. J. Phys. 60, 994-1003 (1992); erratum, 61, 81 (1993).
\bibitem {ST1} David Sokoloff, Ronald Thornton, {\it Using interactive lecture demonstrations to create an active learning environment}, Phys. Teach. 35, 340-347 (1997).
\bibitem {ST2} David Sokoloff, Ronald Thornton, {\it Learning motion concepts using real-time microcomputer-based laboratory tools}, Am. J. Phys. 58, 858-867 (1990).
\bibitem {RSS} E. F. Redish, J. M. Saul, R. N. Steinberg, {\it On the effectiveness of active-engagement microcomputer-based laboratories}, Am. J. Phys. 65, 45-54 (1997).
\bibitem {MRZ}  L. C. McDermott, M. L. Rosenquist, E. H. van Zee, {\it Student difficulties in connecting graphs and physics: Examples from kinematics}, Am. J. Phys. 55, 503-513 (1987).
\bibitem {GA} F. M. Goldberg, J. H. Anderson, {\it Student difficulties with graphical representations of negative values of velocity},  Phys. Teach. 27, 254- 260 (1989).
\bibitem {RJB} R. J. Beichner, {\it Testing student interpretation of kinematics graphs}, Am. J. Phys. 62, 750-762 (1994).

\end{thebibliography}
\end {document}